\documentclass[fleqn,usenatbib]{mnras}
\usepackage{natbib}
\usepackage{amsmath}

\newcommand{\pris}{\emph{Pristine }}
\newcommand{\logg}{\ensuremath{\log g}}

\newcommand{\feh}{\ensuremath{\protect\rm [Fe/H] } }
\newcommand{\car}{\ensuremath{\protect\rm [C/Fe] }}
\newcommand{\teff}{T$_{\rm eff}$}
\newcommand{\SRemp}{23 }
\newcommand{\SRvmp}{56 }
\newcommand{\totstars}{1007 }
\usepackage{graphicx}
\usepackage{tablefootnote}

\usepackage{newtxtext,newtxmath}
\usepackage[T1]{fontenc}
\usepackage{ae,aecompl}

\usepackage{graphicx}	% Including figure files
\usepackage{amsmath}	% Advanced maths commands
\usepackage{amssymb}	% Extra maths symbols

\title[Pristine VI]{The \pris Survey -- VI. The first three years of medium-resolution follow-up spectroscopy of \pris EMP star candidates\thanks{Based on observations made with ESO Telescopes at the La Silla Paranal Observatory under programmes ID 097.B-0764(A) and 0102.B-0449(A); with WHT and INT telescopes at the Observatorio Roque de los Muchachos-- Isla de La Palma under programmes C71, C54, C31, C75, C123, C175, N5, N3, N2, P8, and P2.}}

\author[D. Aguado et al.]{David~S. Aguado$^{1}$\thanks{E-mail: daguado@ast.cam.ac.uk},
Kris Youakim$^{2}$,
Jonay~I. Gonz\'alez~Hern\'andez$^{3,4}$,
\newauthor Carlos Allende Prieto$^{3,4}$,
Else Starkenburg$^{2}$,
Nicolas Martin$^{5,6}$,
Piercarlo Bonifacio$^{7}$,
\newauthor Anke Arentsen$^{2}$,
Elisabetta Caffau$^{7}$, 
Luis Peralta de Arriba$^{1}$,
Federico Sestito$^{5,2}$,
\newauthor Rafael Garcia-Diaz$^{3,4}$,
Nicholas Fantin$^{8}$,
Vanessa Hill$^{9}$,
Pascale Jablonca$^{7,10}$, 
\newauthor Farbod Jahandar$^{11}$,
Collin Kielty$^{12}$,
Nicolas Longeard$^{5}$,
Romain Lucchesi$^{10}$,
\newauthor Rub\'en S\'anchez-Janssen$^{13}$,
Yeisson Osorio$^{3,4}$,
Pedro A. Palicio$^{3,4}$,
Eline Tolstoy$^{14}$,
\newauthor Thomas G. Wilson$^{15,16}$,
Patrick C\^ot\'e$^{8}$,
Georges Kordopatis$^{9}$,
Carmela Lardo$^{10}$,
\newauthor Julio F. Navarro$^{12}$,
Guillaume F. Thomas$^{8}$,
Kim Venn$^{12}$
\\
\\
$^{1}$Institute of Astronomy, University of Cambridge, Madingley Road, Cambridge CB3 0HA, UK \\
$^{2}$Leibniz-Institut fur Astrophysik Potsdam, An der Sternwarte 16, D-14482 Potsdam, Germany\\
$^{3}$Instituto de Astrof\'{\i}sica de Canarias,
              V\'{\i}a L\'actea, 38205 La Laguna, Tenerife, Spain\\
$^{4}$Universidad de La Laguna, Departamento de Astrof\'{\i}sica, 
             38206 La Laguna, Tenerife, Spain \\
$^{5}$Universit\'e de Strasbourg, CNRS, Observatoire astronomique de Strasbourg, UMR 7550, F-67000 Strasbourg, France\\
$^{6}$Max-Planck-Institut f\"{u}r Astronomie, K\"{o}nigstuhl 17, D-69117 Heidelberg, Germany \\
$^{7}$GEPI, Observatoire de Paris, Universit\'e PSL, CNRS, Place Jules Janssen, 92190, Meudon, France \\
$^{8}$NRC Herzberg Astronomy and Astrophysics, 5071 West Saanich Road, Victoria, BC, V9E 2E7, Canada\\
$^{9}$Universite C\^ote d'Azur, Observatoire de la C\^ote d'Azur, CNRS, Lagrange, Bd de l'Observatoire, CS34229, F-06304 Nice, France\\
$^{10}$Institute of Physics, Laboratoire d'astrophysique, Ecole Polytechnique F\'ed\'erale de Lausanne (EPFL), Observatoire, 1290 Versoix, Switzerland\\
$^{11}$University of Victoria, 3800 Finnerty Rd, Victoria, BC, V8P 5C2, Canada\\
$^{12}$Department of Physics and Astronomy, University of Victoria, P.O. Box 3055, STN CSC, Victoria BC V8W 3P6, Canada\\
$^{13}$UK Astronomy Technology Centre, Royal Observatory Edinburgh, Blackford Hill, Edinburgh, EH9 3HJ, UK\\
$^{14}$Kapteyn Astronomical Institute, University of Groningen, Landleven 12,NL-9747 AD Groningen, the Netherlands\\
$^{15}$Department of Physics \& Astronomy, University College London, London WC1E 6BT, UK\\
$^{16}$Isaac Newton Group, Apartado 321, E-38700 Santa Cruz de La Palma, Spain\\
}
\date{Accepted XXX. Received YYY; in original form ZZZ}

\pubyear{2019}

\begin{document}
\label{firstpage}
\pagerange{\pageref{firstpage}--\pageref{lastpage}}
\maketitle

\begin{abstract}
We present the results of a 3-year long, medium-resolution spectroscopic campaign aimed at identifying very metal-poor stars from candidates selected with the $CaHK$, metallicity-sensitive Pristine survey. The catalogue consists of a total of \totstars stars, and includes 146 rediscoveries of metal-poor stars already presented in previous surveys, 707 new very metal-poor stars with $\feh < -2.0$, and 95 new extremely metal-poor stars with $\feh < -3.0$. We provide a spectroscopic [Fe/H] for every star in the catalogue, and [C/Fe] measurements for a subset of the stars (10\% with $\feh < -3$ and 24\% with $-3 < \feh < -2$) for which a carbon determination is possible, contingent mainly on the carbon abundance, effective temperature and S/N of the stellar spectra. We find an average carbon enhancement fraction ($[C/Fe] \geq +0.7$) of 41 $\pm$ 4\% for stars with $-3 < \feh < -2$ and 58 $\pm$ 14\% for stars with $\feh < -3$, and report updated success rates for the \pris survey of \SRvmp\% and \SRemp\% to recover stars with $\feh < -2.5$ and $\feh < -3$, respectively. Finally, we discuss the current status of the survey and its preparation for providing targets to upcoming multi-object spectroscopic surveys such as WEAVE.   
\end{abstract}


%%%%%%%%%%%%%%%%%%%%%%%%%%%%%%%%%%%%%%%%%%%%%%%%%%

%%%%%%%%%%%%%%%%% BODY OF PAPER %%%%%%%%%%%%%%%%%%

\section{Introduction}
The current picture of Galactic chemical enrichment is based on the production of elements heavier than He in the interiors of stars, their subsequent release into the interstellar medium through supernova explosions, and their eventual reintegration into ensuing stellar generations. Apart from a few exceptions, such as mass transfer binaries, the current elemental compositions of stars are expected to maintain the chemical imprint of their birth environments, which in turn reflect this enrichment process. Based on this principle, it is possible to use stars with primitive elemental abundance patterns, also known as very metal-poor (VMP: \feh$<-2$), to study the early Universe. 

%A wealth of information is contained in these stars, including their locations, orbits, and detailed chemical abundance signatures, all of which shed light on the specific processes that led to their formation, including the properties of previous stellar generations, all the way back to the very first stars that formed in the Universe. 

One issue that hampers our ability to study the detailed abundance trends of metal-poor stars, is their scarcity in our local environment with respect to the younger, more metal-rich populations. However, metal-poor stars are more abundant in certain Galactic environments, making them promising searching grounds. 
Cosmological simulations demonstrate that the outer regions of the Galaxy are the most dominated by old and/or metal-poor stars (see for recent studies using hydrodynamical simulations \citet{sta17II} \& \citet{elbad18}). If one has a good method to efficiently distinguish metal-poor from more metal-rich populations and is interested in the oldest stars among the most metal-poor, then the Galaxy's inner regions and some of its satellites are also promising hunting grounds \citep[e.g.,][]{white00,tum10,sta17II}.
%Springel & White 2000, Tumlinson 2010, Starkenburg et al., 2017)

%Using the FIRE cosmological zoom-in simulations of three Milky Way-mass disk galaxies, \citet{elbad18} suggest that the highest densities of old stars should be found on lines of sight towards the outer galaxy. Another study by \citet{sta17II} using Milky Way-sized galaxies selected from the APOSTLE cosmological hydrodynamical simulations confirms that the fractions of metal-poor stars and old stars increase at larger Galactic radii, and are also relatively high in satellite dwarf galaxies. They also note that the fraction of stars that are both metal-poor and old is highest in the Galactic centre, and suggest that this is also a promising region to observationally search for old, metal-poor stars. 

%Different Galactic environments host different chemical populations of stars 
\begin{figure*}
	\includegraphics[width=\textwidth]{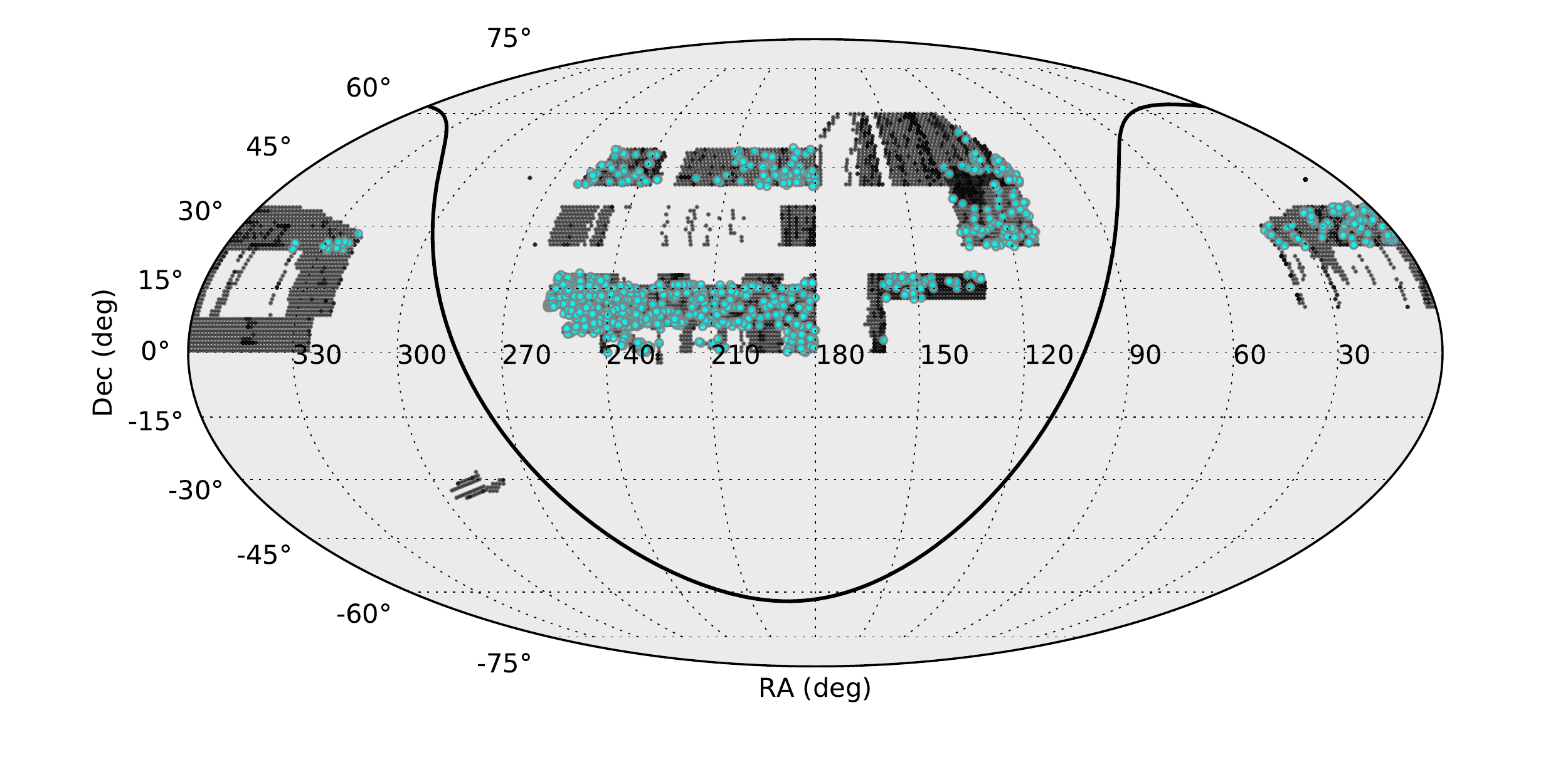}
        \caption{The current footprint of the \pris survey (black) covering $\sim 5000\, \mathrm{deg}^2$. The stars making up the spectroscopic follow-up sample are plotted as cyan points and were selected from $\sim 2500\, \mathrm{deg}^2$ of the total region. The Galactic plane is shown as the black line.}
    \label{footprint}
\end{figure*}

\begin{table*}
\begin{center}
\renewcommand{\tabcolsep}{5pt}
\centering
\caption{Technical Information for Facilities Used in this Analysis}
\label{techni}
\begin{tabular}{ccccccccc}
\hline
\multicolumn{1}{c}{Instrument}&\multicolumn{1}{c}{Telescope}&\multicolumn{1}{c}{Detector} & \multicolumn{1}{c}{Grating}& \multicolumn{1}{c}{Dispersion}& \multicolumn{1}{c}{Range} & \multicolumn{1}{c} {Resolution at $\sim 4500$\AA}& \multicolumn{1}{c} {Slit} \\
\hline

$\rm IDS$      & 2.5-m INT &EEV10 &R900V  &0.69\,\AA\,px$^{-1}$ & 3600-5200\,\AA &   3300 & $1\farcs0$   \\
$\rm ISIS$     & 4.2-m WHT &EEV12 &R600B  &0.45\,\AA\,px$^{-1}$ & 3600-5100\,\AA &   2400 & $1\farcs0$     \\
$\rm EFOSC2$   & 3.6-m NTT &CCD40 &600    &0.95\,\AA\,px$^{-1}$ & 3600-5200\,\AA &   930 & $0\farcs7-1\farcs0$      \\
\hline
\end{tabular}
\end{center}
\end{table*}

Naturally, a substantial amount of effort has gone into finding and studying these rare stars, and they remain a strong focus of current and future surveys dedicated to Galactic Archaeology. Given that they are so rare among the far more numerous foreground populations, there are two options when searching for metal-poor stars: (i)observing a large sample of stars from general science purpose surveys to find the few metal-poor stars among them, or (ii)targeted searches which aim for these stars specifically. The former approach has been quite successful and has contributed significantly to the current sample of the most metal-poor stars \citep[e.g.][]{caff13I,aoki13,alle15,agu16,li15,agu17II,agu17I,agu18II,agu18I}, mostly with the help of large spectroscopic surveys such as the Sloan Digital Sky Survey \citep[SDSS,][]{yor00}, the Sloan Extension for Galactic Understanding and
Exploration (SEGUE, \citealt{yan09}), the Baryonic Oscillations 
Spectroscopic Survey (BOSS, \citealt{eis11,daw13}), and more recently the  large Sky Area 
Multi-Object Fiber Spectroscopic Telescope (LAMOST, \citealt{deng12}). More targeted searches have also been in use for many years, from early efforts using a Ca H \& K objective-prism technique, such as the HK survey \citep{bee85,bee92} and the Hamburg ESO survey \citep{chris02}, to more recent efforts using targeted narow/medium-band photometry at blue wavelengths, like the \emph{SkyMapper} survey \citep{kell07,wolf18,casa19,hua19}, and the \pris survey \citep{sta17I}.

Future metal-poor star searches will be even more effective by combining both of these strategies. The upcoming generation of multi-object spectroscopic (MOS) surveys such as the William Herschel Telescope Enhanced Area Velocity Explorer \citep[WEAVE]{dal18}, the Dark Energy Spectroscopic Instrument \citep[DESI]{levi13}, the 4-metre Multi-Object Spectroscopic Telescope \citep[4MOST]{jong19}, the Galactic Archaeology with HERMES \citep[GALAH]{bud18}, the Sloan Sky Digital Survey-V \citep[SDSS-V]{col17}, and the Maunakea Spectroscopic Explorer \citep[MSE]{mac16,mac19} will provide, together with Gaia \citep{gaia2018}, an unprecedented number of spectra over the whole sky. Although these surveys will have the capability to observe tens of millions of stars, it will still be necessary to target metal-poor stars specifically in order to maximize the output for Galactic Archaeology studies. When used in tandem with pre-selection surveys such as \emph{SkyMapper}, and \emph{Pristine}, it will be possible to obtain high quality observations of metal-poor stars across an unprecedented range of magnitudes, wavelengths, and Galactic environments. In order for the target pre-selection from such surveys to be maximally effective, they must be validated beforehand by dedicated spectroscopic follow-up programs.

In this paper, we present the results of the first three years of spectroscopic follow-up for the \pris survey, using low- and medium-resolution spectroscopic facilities. This not only provides a detailed understanding of the selection of candidates to target with future MOS surveys, but also has the added value of providing the Galactic Archaeology community with a sizeable catalogue of new, metal-poor stars, a subset of which also have measurements of carbon abundances.

Carbon abundance is a well studied quantity in metal-poor stars, and has important implications for understanding the earliest stellar generations. First, the carbon abundance of a star influences the cooling channels and may allow for low-mass star formation \citep{brom03}. Secondly, as discussed at length in \citet{bee05,yong13II,boni15,yoo16,chi19}, the increase in carbon-enhancement with decreasing metallicity in EMP stars allow us to make a phenomenological taxonomy of ancient stars. 

There are two definitions for carbon-enhanced metal-poor (CEMP) stars currently presented in the literature. \citet{bee05} propose a definition of CEMP stars as stars with $\car>+1.0$\footnote{More recently \citet{boni18} proposed to establish a fixed $\rm A(C)>5.5$ reference value for stars with $\feh<-4.0$ to be CEMP stars.}, while \citet{aoki07} use $\car>+0.7$ with an additional correction depending on the luminosity.
These different values do not reflect theoretical studies but still provide a useful quantitative classification. On the other hand, the original critical carbon abundance from \citet{bro03} ($[\rm C/H]_{crit}\simeq-3.5\pm 0.1$) has recently been improved to include the effect of the silicate grains in cooling processes allowing for fragmentation of the proto-stellar clouds \citep{chi17}. These studies propose three regions in the $\rm A(C)-[Fe/H]$ plane: the carbon dominated, the silicate dominated area and the forbidden area due the insufficient dust cooling. So far only one star, J1029+1729, belonging to the \feh$<-4.5$ regime is clearly carbon-normal \citep{caff11} with $\car<+0.7$. J1029+1729 is still the most metal-poor star known but remains in the silicate dominated region well below $\car=+2.3$ line. Discovered by \citet{sta18} and included in this work, $Pristine\,221.8781+9.7844$ is the second most metal-poor star also in the silicate dominated region with $\car<+1.76$ and could also potentially be a carbon-normal UMP star. All 11 other stars from the literature with \feh < -4.5 show a clear enhancement in carbon \cite[see, e.g.,][and references therein]{boni18,yoon19}. Larger samples of extremely metal-poor stars, especially those with robust carbon measurements, are important in order to better understand these trends.

The paper is organized as follows. In Section \ref{obs}, we summarize the data set, observations, and reduction methods. In Section \ref{analysis}, the analysis of the data using the FERRE code is explained. In section \ref{catalogue}, we present the spectroscopic follow-up catalogue, including a discussion of the updated success rates for finding EMP and VMP stars of the \pris survey. In Section \ref{future}, we look at the future of \pris and its synergies with other upcoming surveys, and we conclude the paper in Section \ref{conclusions}.
\begin{table*}
\begin{center}
\caption{FERRE analysis for a sample of well known EMP stars. Uncertainties include both systematic and statistical errors. Values from the literature derived from high-resolution analyses are also shown.}
\label{emp_table}
\begin{tabular}{lccccccccc}
\hline
&\multicolumn{3}{c}{Values from FERRE} &  &  & \multicolumn{4}{c}{Values from the literature} \\
\cline{2-4}  \cline{7-10}
Object & \teff\,[K] & \logg& \feh& S/N & Inst.  &\teff\,[K] & \logg& \feh& Ref. \\
\hline
 HE 0057$-$5959    & $5333\pm118$ & $1.72\pm0.72$& $-3.69\pm0.27$ &39  &EFOSC &$5257\pm100$ &$1.72\pm0.30$  &$-4.08\pm0.30$ & 1 \\
 SDSS J0723$+$3637 & $5258\pm212$ & $2.70\pm1.33$& $-3.41\pm0.21$ &28  &IDS   &$5150\pm150$ &$2.20\pm0.50 $  &$-3.32\pm0.20$  & 2 \\
  HD 84937         & $6379\pm109$ & $4.75\pm0.50$& $-2.19\pm0.21$ &181 &IDS   &$6431\pm100$ &$4.08\pm0.30$  &$-2.14\pm0.20$  & 3 \\
SDSS J1004$+$3442  & $6002\pm140$ & $2.84\pm0.95$& $-2.83\pm0.25$ &13  &IDS   &$6100\pm150$ &$4.00\pm0.50 $  &$-3.09\pm0.20$  & 2 \\
 SDSS J1036$+$1212 & $6052\pm102$ & $1.26\pm0.50$& $-3.24\pm0.21$ &34  &IDS   &$5850\pm150$ &$4.00\pm0.50 $  &$-3.47\pm0.20$  & 2 \\
 SDSS J1108$+$1747 & $5930\pm104$ & $4.89\pm0.50$& $-3.07\pm0.21$ &35  &IDS   &$6050\pm150$ &$4.00\pm0.50 $  &$-3.17\pm0.20$  & 2  \\
 SDSS J1128$+$3841 & $6416\pm126$ & $4.61\pm0.61$& $-3.28\pm0.22$ &39  &IDS   &$6550\pm150$ &$4.00\pm0.50 $  &$-2.82\pm0.20$  & 2 \\
 HE 1207$-$3108    & $5545\pm156$ & $3.11\pm0.87$& $-3.01\pm0.22$ &93  &EFOSC &$5294\pm100$ &$2.85\pm0.30$  &$-2.70\pm0.30$  & 1 \\
 HE 1320$-$2952    & $5658\pm123$ & $4.09\pm0.59$& $-3.13\pm0.22$ &50  &EFOSC &$5106\pm100$ &$2.26\pm0.30$  &$-3.69\pm0.30$ & 1 \\
 HE 1327$-$2326    & $6400\pm109$ & $4.82\pm0.50$& $-5.40\pm0.43$ &30  &IDS   &$6180\pm80$  &$4.50 \pm0.50 $  &$-5.70\pm0.20 $ & 4 \\
  G64$-$12         & $6435\pm105$ & $4.97\pm0.50$& $-3.24\pm0.22$ &80  &IDS   &$6550\pm100$ &$4.68\pm0.30$  &$-3.21\pm0.20$  & 3 \\
 CS 30336$-$0049   & $5194\pm161$ & $2.60\pm1.14$& $-3.97\pm0.22$ &51  &EFOSC &$4725\pm100$ &$1.19\pm0.30$  &$-4.10\pm0.30$  & 1 \\
HE 2047$-$5612     & $6281\pm122$ & $4.64\pm0.55$& $-2.94\pm0.22$ &41  &EFOSC &$6128\pm100$ &$3.68\pm0.30$  &$-3.14\pm0.30$  & 1 \\
 SDSS J2206$-$0925 & $5210\pm100$ & $1.01\pm0.50$& $-2.66\pm0.20$ &29  &IDS   &$5100\pm150$ &$2.10\pm0.50$  &$-3.17\pm0.20$  & 2 \\
 BD+17 4708        & $6100\pm106$ & $3.90\pm0.50$& $-1.80\pm0.21$ &120 &IDS   &$6085\pm50$ &$4.10\pm0.10$  &$-1.60\pm0.10 $ &5  \\
 SDSS J2338$-$0902 & $5052\pm101$ & $1.03\pm0.50$& $-2.62\pm0.20$ &32  &IDS   &$4900\pm150$ &$1.90\pm0.50 $  &$-3.12\pm0.20$  & 2 \\
\hline
\end{tabular}
References: 1=\citet{yong13II}; 2=\citet{aoki13}; 3=\citet{ishi12}; 4=\citet{fre05}; 5=\citet{gra03}
\end{center}
\end{table*}

\section{Data and Observations}
\label{obs}
As discussed in detail in \citet{sta17I}, one of the main aims of the \pris project is to enlarge the number of metal-poor stars currently known in our Galaxy and characterize them to better understand the Galactic halo. Figure \ref{footprint} shows the current \pris footprint which covers a total of $\sim 5000\textrm{ deg}^2$ in the Northern Galactic halo. The targets selected for follow-up spectroscopy are shown in cyan, and were selected form a $\sim2500\textrm{ deg}^2$ region of the total footprint. 

\begin{figure}
\label{v_mag_hist}
	\includegraphics[width=\columnwidth]{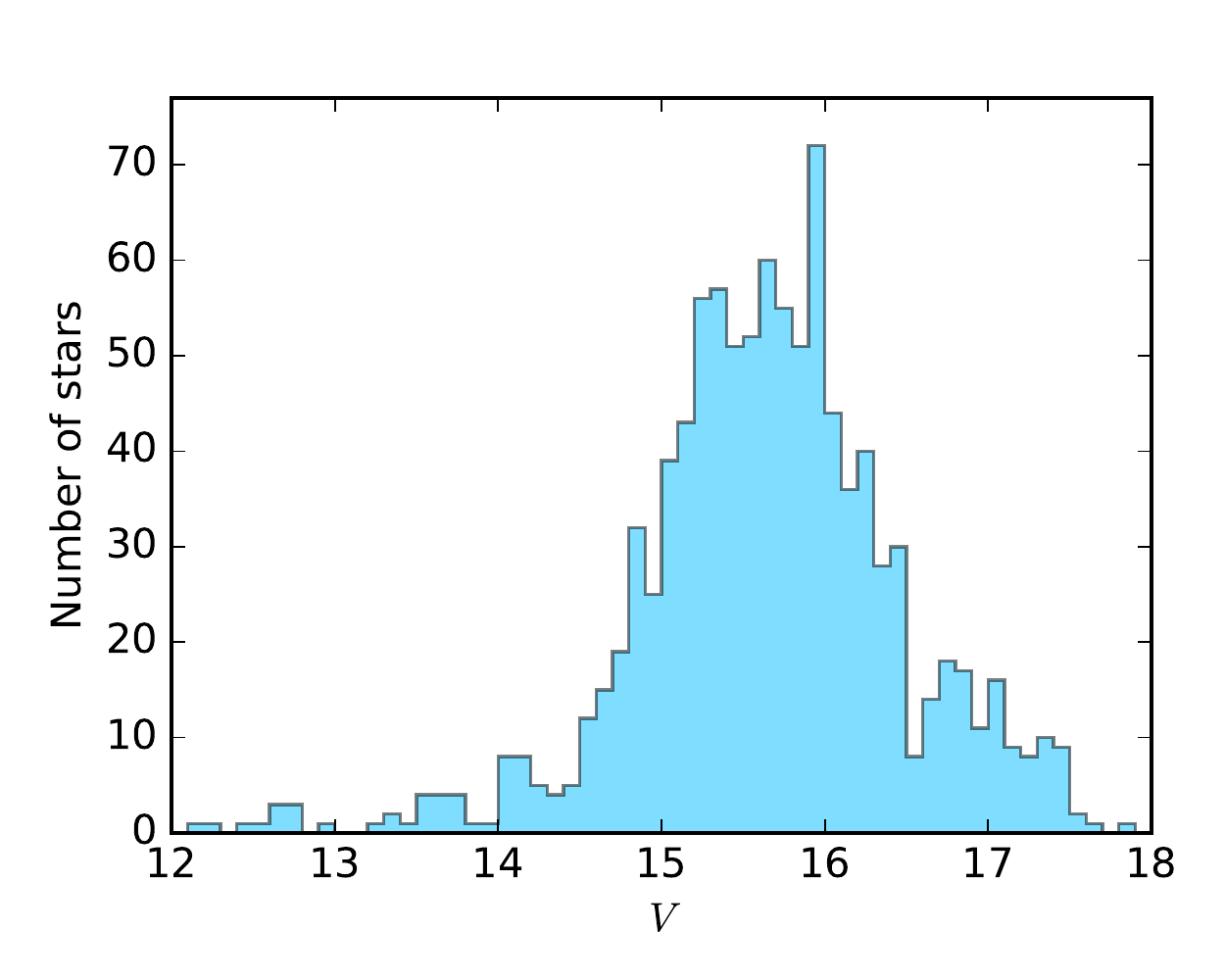}
        \caption{Distribution of $V$ magnitudes for the full follow-up spectroscopic sample of \totstars \pris stars.}
    \label{V_mag_hist}
\end{figure}

\subsection{Observations}
The spectroscopic data presented here were collected over a period of 6 semesters, from March 2016 to February 2019. Figure \ref{v_mag_hist} shows the V-band magnitude distribution of the spectroscopic follow-up sample, totalling 1008 stars. Due to the wide range in target brightness, three different facilities were used to conduct follow-up observations of EMP candidates selected from the \pris survey: the Intermediate Dispersion Spectrograph (IDS) on the 2.5-m Isaac Newton Telescope (INT), the Intermediate-dispersion Spectrograph and Imaging System \citep[ISIS,][]{isiswht} on the 4.2-m William Herschel Telescope, and the ESO Faint Object Spectrograph and Camera \citep[EFOSC2,][]{efosc84} on the 3.6-m New Technology Telescope (NTT). The selected mode in all cases was \textit{long slit} providing low- and medium-resolution spectroscopy (see Table \ref{techni} for further technical details).

Fainter targets  (g $>16.2$) were observed with the larger aperture WHT and NTT telescopes, while brighter targets (g $<16.2$) were observed with the INT. The total number of observing nights were 182 (145 with IDS, 25 with ISIS and 12 with EFOSC). Although the ISIS observations were shared with another program so that the resulting equivalent observing nights came out to $\sim$10.
\begin{figure*}
	\includegraphics[width=150mm]{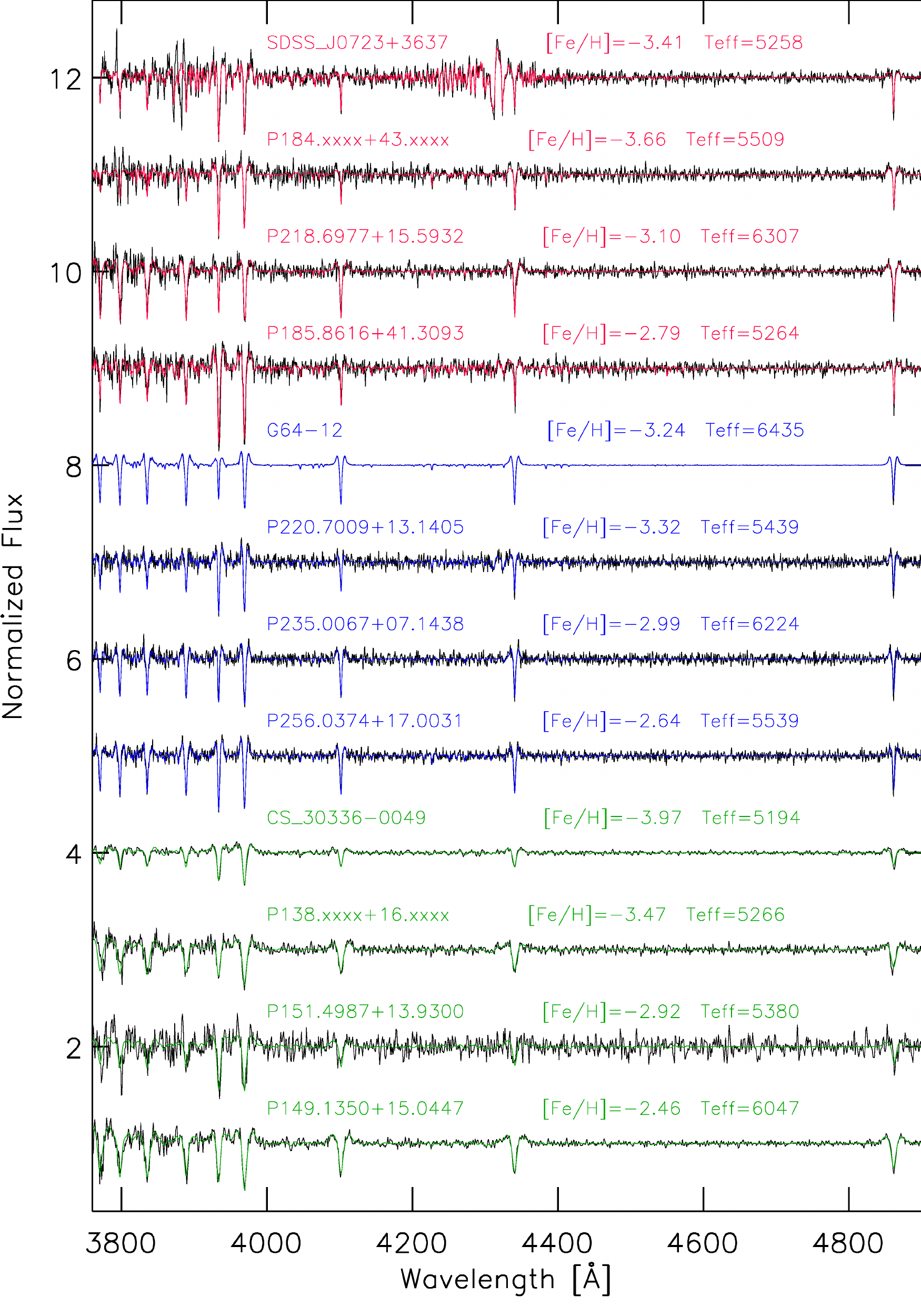}
        \caption{A subsample of the spectra of the \pris observed targets with IDS/INT (red), ISIS/WHT (blue), and EFOSC/3.6\,NNT (Green) together with the best fit derived with FERRE. Three well known metal-poor stars are shown for comparison, SDSS J0723+3637, G64$-$12, and CS 30336$-$0049. The main stellar parameters are also displayed. }
    \label{temp_spec}
\end{figure*}
\subsection{Observational strategy}
The minimum desired signal-to-noise (S/N) ratio per pixel for the observations was $\sim15-25$ in the calcium H \& K spectral region ($\sim3950$\,\AA), depending on the effective temperature of the specific star. Therefore, the average exposure time for a single integration was 1500\,s,  900\,s and 1500\,s, for the INT, ISIS and EFOSC observations, respectively. Naturally, exposure times varied slightly for each individual object depending on the target brightness and the visibility conditions. 
The observational strategy was designed to maximize the ratio between the number of observed candidates and the reliability of the derived parameters. However, stars that were identified as ultra metal-poor (UMP) candidates during an observing run were subsequently followed-up with more exposures to achieve a higher S/N. Stars that still seemed highly interesting at this stage were then followed up with larger telescopes at higher resolution. This observing strategy was designed to maximize the detection of very low-metallicity stars, and has yielded the discovery of $Pristine\,221.8781+9.7844$, an ultra metal-poor sub-giant star with $\feh= -4.66\pm0.13$ and [C/Fe]$ < 1.76$. The detailed analysis of this star with high resolution follow-up with VLT/UVES is described in \citet{sta18}.

%\subsection{Spectroscopic follow-Up}
\subsection{Data reduction}
The spectral data reduction included bias substraction, flat-fielding, and wavelength calibration -- using CuNe$+$CuAr lamps for IDS and ISIS, and He$+$Ar for EFOSC--, and was performed using the \textsc{onespec} package in \textsc{IRAF} \citep{tod93}. At the moderate S/N levels required for this program and at medium-resolution, the contribution of the interstellar medium (ISM) in the Ca H \& K area is, in general, not resolved \citep[see e.g.,][]{agu16,agu17II}. In order to reduce the uncertainties from the spectral analysis, we remove the bluest part of the spectrum most affected by noise, considering only the region redder than 3700\,\AA. 

\section{Analysis with FERRE}
\label{analysis}

\begin{figure}
	\includegraphics[width=\columnwidth]{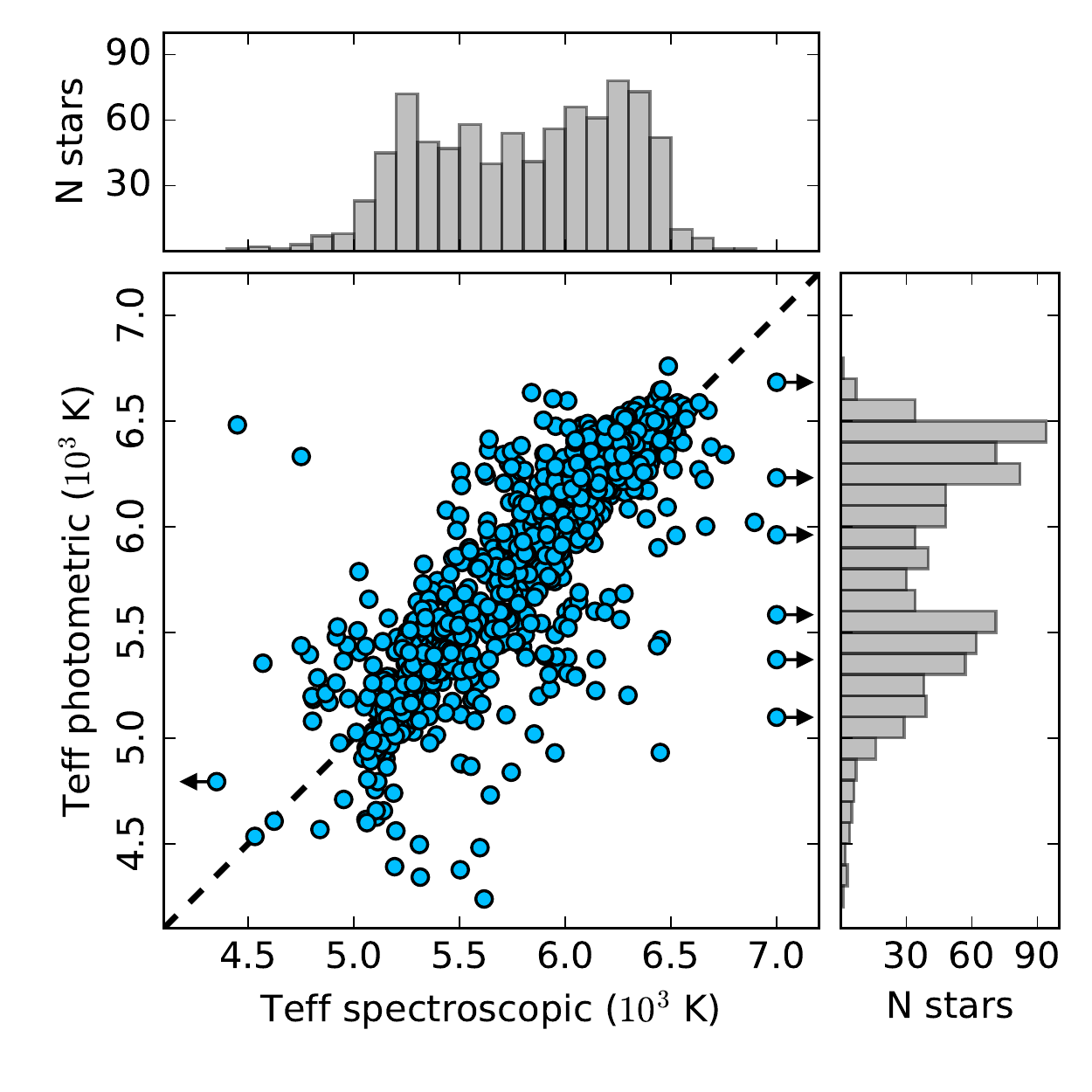}
        \caption{Comparison of the photometric temperatures to those derived spectroscopically with FERRE. Stars that fall outside of the plotted region are plotted on the edge, and marked with an arrow. Histograms of both distributions are shown to indicate the density of the points.}
    \label{temp_plot}
\end{figure}

The entire sample of the spectroscopic data has been analyzed using the grid of synthetic stellar spectra computed with the ASSET code \citep{koe08} and published in \citet[][hereafter DA17]{agu17I}. The model atmospheres were computed with the Kurucz codes, and are described in \citet{mez12}. We use the FERRE\footnote{{\tt FERRE} is available from http://github.com/callendeprieto/ferre} code \citep{alle06} to search for the best fit to the observed spectrum by simultaneously deriving the main three stellar atmospheric parameters (effective temperature \teff, surface gravity \logg, metallicity [Fe/H]), and carbon abundance [C/Fe]. FERRE is able to interpolate between the nodes of the grid and provide a synthetic spectrum for each set of derived parameters. A similar analysis in implemented in \citet[][hereafter KY17]{kris17} used the Powel's truncated Newton algorithm to find the best fit solution. However, for this work we also use a Markov Chain Monte Carlo (MCMC) algorithm based on self-adaptative randomized subspace sampling \citep{vru09}, which provides the added advantage of deriving uncertainties by sampling the probability distribution function. The grid of synthetic spectra spans the space $-6\leq[\rm Fe/H]\leq-2$, $-1 \leq[\rm C/Fe]\leq 5$, $4750\,\rm K\leq T_{eff}\leq7000\,\rm K$ and $1.0\leq \logg \leq5.0$. Although we targeted objects in the $\feh_{Pristine}<-2$ regime, there were some stars that were observed with higher metallicities. Those targets were re-analyzed with a more generic grid, suitable for higher metallicities, and described in \citet{alle18}. 

In order to cross-validate our analysis methods, we observed a number of well-known EMP stars from the literature that have robust stellar parameter determinations from high-resolution analyses. Comparing those stellar parameters with the ones measured in this work, we find a median deviation of 177\,K, 0.86, and 0.27\,dex for \teff, \logg, and \feh, respectively. Table \ref{emp_table} summarizes the FERRE analysis performed on this sample and demonstrates that our derived metallicities are in very good agreement with those from the literature, thus demonstrating the ability of our method to derive precise metallicities using medium-resolution spectra. More comparisons of stellar parameter determination with the FERRE code and standard stars in the literature can be found in DA17. Figure \ref{temp_spec} shows a subsample of the observed spectra together with the best fit synthetic spectrum as determined by FERRE for each of the three different instruments as well as three more well-known metal-poor stars.

\subsection{Stellar parameters}
\label{stellar_params}

To simultaneously derive \teff, \logg, \feh, and [C/Fe], we smooth the grid of models and resample them to the appropriate resolving power corresponding to each instrument (see Table \ref{techni}). We then normalize both the synthetic models and the observed spectra using a running-mean filter with a 30-pixel window (see DA17 for further details). Finally, FERRE derives the set of parameters assuming $[\alpha/\rm Fe]=+0.4$ and a fixed value of the microturbulence of 2.0\,km s$^{-1}$.

\teff\, is obtained by fitting the entire spectrum, although the derived \teff\ is largely influenced by the Balmer lines present in the spectral range ($\rm H_{\beta}$-4861\AA, $\rm H_{\gamma}$-4340\AA, $\rm H_{\delta}$-4101\AA, $\rm H_{\epsilon}$-3970\AA, $\rm H_{\zeta}$-3889\AA, $\rm H_{\eta}$-3835\AA, $\rm H_{\theta}$-3797\AA, $\rm H_{\iota}$-3770\AA, $\rm H_{\kappa}$-3750\AA, $\rm H_{\lambda}$-3734\AA, $\rm H_{\mu}$-3721\AA). The temperature determination method relies on the broadening theory of the Balmer lines which is described in \citet{ber00d}. The running mean normalization reduces the dependence on the specific determination of the continuum, allowing improved temperature determinations based on the shape of each H line, even with a moderate S/N ($\sim15-20$). DA17 consider a systematic uncertainty for deriving temperatures of $\delta$\teff$=100$\,K,\, which is then combined quadratically with the statistical error from the MCMC method. Referring back to Table \ref{emp_table}, the derived effective temperatures are fully compatible with those from previous works. In Figure \ref{temp_plot}, we show the relation between the photometric temperatures derived using the SDSS $(g-i)$-temperature relation\footnote{For the equation used to compute the photometric temperatures, see the InfraRed Flux Method (IRFM), \url{https://www.sdss.org/dr12/spectro/sspp_irfm/}, [Fe/H] = -2.5 was asssumed.} and the temperatures derived from FERRE using the spectroscopic data from IDS, ISIS, and EFOSC. 

 Measuring \logg\, values from medium-resolution spectra when no \ion{Fe}{ii} lines are available is a challenge. Particularly at moderate S/N ($\sim15-25$), the shape of the Balmer lines alone do not allow for it to be derived precisely. However, a coarse classification between the dwarf/giant regimes is possible with FERRE. Robust \logg\, determination in metal-poor stars using Gaia data is possible, but good quality parallax measurements are required. Since this is not available for most of our sample, particularly the fainter objects, we use the spectroscopic values from FERRE and assume the same systematic error as were used in DA17 of $\delta$\logg$=0.5$. 
 
\begin{figure}
\begin{center}
	\includegraphics[width=75mm]{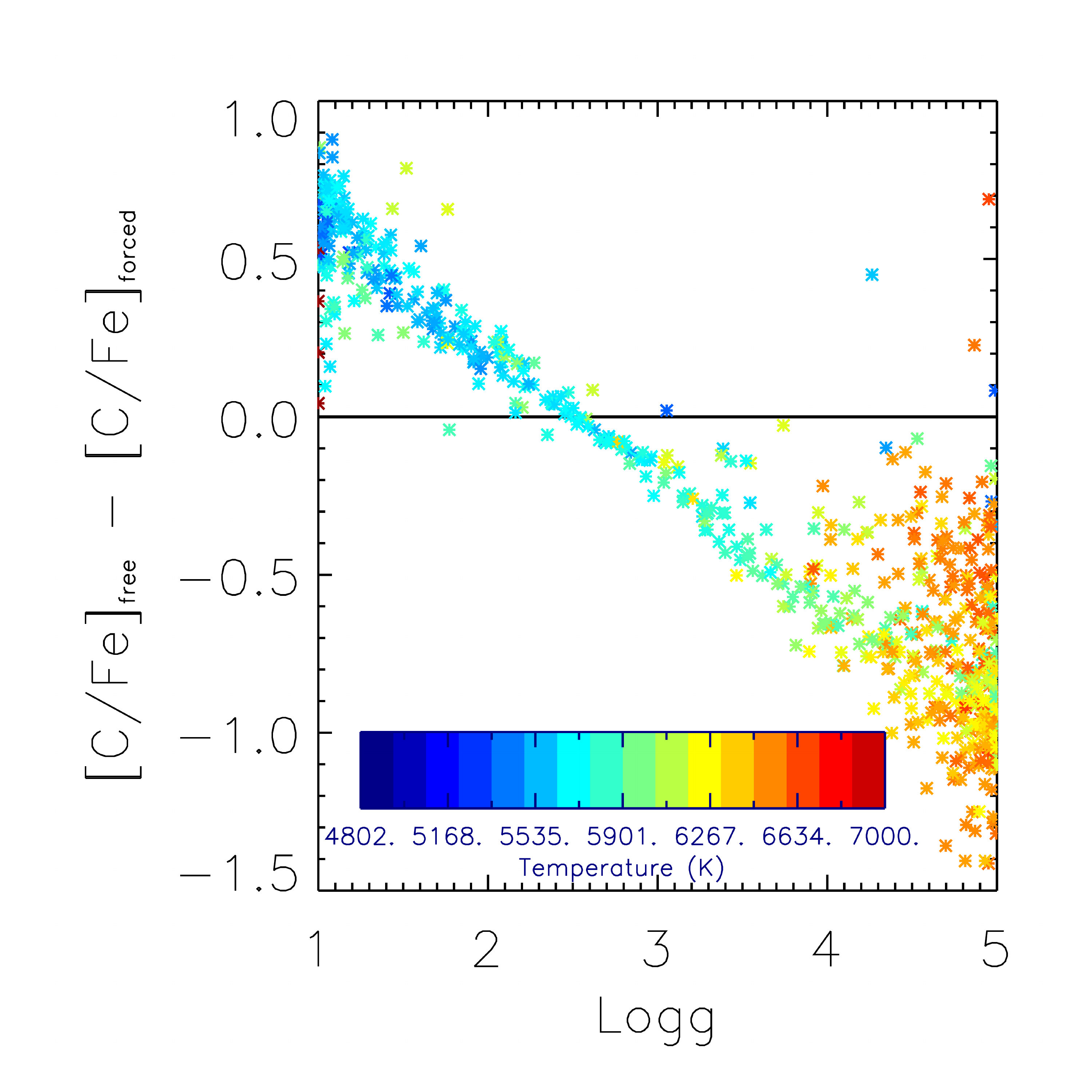}
	\caption{An analysis over the \textit{Pristine} subsample observed with IDS on the INT. The vertical axis show the difference between the original \car\, determination and the one we derive assuming a fixed $\logg=2.5$ versus the derived \logg\, colour coded by \teff.}
    \label{gravity_plot}
\end{center}
\end{figure}
	
The deepest metallic absorption in the optical range is caused by the calcium H\&K resonant lines at 3933\,\AA\, and 3968\,\AA\, respectively. Using these features as robust indicators to infer metallicities in EMP stars with low/med-resolution spectroscopy is a longstanding method \citep[see e.g.][]{bee85,bee92,rya91,car96}, and is still used today in large spectroscopic surveys such as SEGUE, BOSS and LAMOST \citep[see e.g.][and references therein]{caff13I, agu16,agu17II,li15,li18,fran18}.  
However, there is additional information present in this spectral range, such as the \ion{Mg}{i}b triplet and some weak \ion{Fe}{i} and \ion{Sr}{ii} lines, and these features can also contribute to the derivation of metallicities, provided that the S/N is high enough to resolve them. Reassuringly, we find good agreement between our [Fe/H] values and those from high-resolution analyses even with the relatively low-resolution EFOSC2 instrument (R$\sim$1000, see Table \ref{emp_table} and section 4.1 in DA17).
DA17 assumed a systematic uncertainty in metallicity of 0.1\,dex. However, due to the significantly lower S/N of the current sample, the ISM contribution to the Ca H \& K absorption lines is largely unresolved for most of the spectra. Therefore, we assume a more conservative value of $\delta\feh=0.2$\,dex and add it to the derived statistical uncertainty. 

%In Section \ref{success_rates}, we present a more detailed analysis of the success rates of the \pris photometric pre-selection of metal-poor stars.

%For a subsample of the \pris targets we were also able to derive carbon abundances. The details of this analyis are presented in the next section.

\begin{figure}
\begin{center}
	\includegraphics[width=85mm]{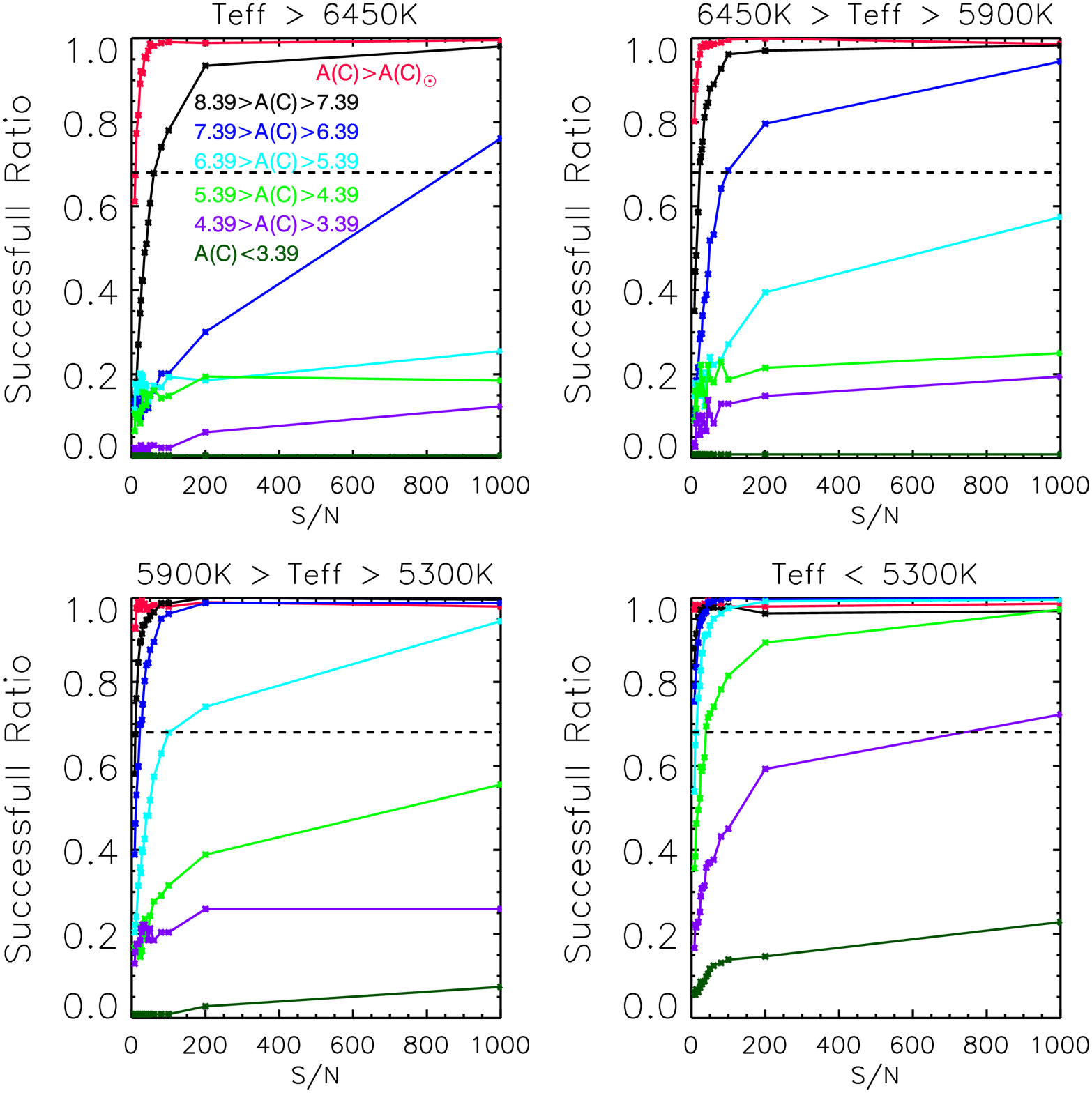}
	\caption{The successful ratio measuring carbon abundance in a set of 30~854 synthetic spectra covering a wide range of effective temperatures, absolute carbon abundance and different levels of signal-to-noise ratio. Dashed line is 68\% of successful ratio.}
    \label{carbon_plot1}
\end{center}
\end{figure}
	
\subsection{Carbon abundance}\label{carbon_sec}
Due to the lack of spectral features in EMP stars, particularly at higher \teff, it is not always possible to derive a reliable carbon abundance, particularly with medium-resolution spectroscopy \citep{boni15}. DA17 provide some reference levels regarding our ability to measure carbon, but as previously discussed, the average S/N in the current sample is significantly lower. With the aim to constrain the confidence levels with which it is possible to derive carbon abundances without important systematic effects, we performed the following theoretical exercise.

A set of synthetic spectra with the same coverage as our IDS/ISIS/EFOSC data (3600-5200\,\AA) were computed with ASSET R$=3000$. A total number of 5670 spectra covering different ranges of \teff, \logg, and absolute carbon abundance, A(C), were analyzed using 10 Markov Chains of 1000 experiments each for different values of S/N ranging from 8 to 200. In total, 30~854 spectra were analyzed with FERRE. We then compared the synthetic absolute carbon abundance A(C) and the corresponding \car\, derived value. We marked a given trial as successful if it was able to recover the theoretical value provided by the synthetic grid, where $\lvert A(C)_{in}-A(C)_{out}\rvert<\epsilon_{A(C)}$, with $\epsilon_{A(C)}$ the assumed systematic uncertainty of 0.2\,dex as estimated in DA17. Figure \ref{carbon_plot1} shows all the ratios versus S/N for different effective temperatures and carbon abundances. For this work, we consider the reliable areas of the plot to be those where the correct value is recovered with a frequency that is higher than 68\%. 
For example, at solar \citep[$A(C)=8.39$,][]{asp05} or higher carbon abundance (red line), we are able to measure \car\, at any temperature with S/N > 10, while for values below $A(C)=4.4$ it is not likely to be able to detect carbon at this resolution, regardless of the \teff. Table \ref{tab_car} summarizes the approximate S/N required to detect the G-band in \pris spectroscopic data. As expected, lower temperatures allow for a better carbon determination due the larger absorption of the G-band. We apply these cuts to the sample and only provide carbon abundance values for the 169 (i.e. 18\%) stars that satisfy these criteria. We note that we are able to measure carbon in 10\% of the stars with \feh$<-3$, and 24\% of the stars with -3 < \feh <-2.

\begin{table}
\begin{center}
\caption{The minimum S/N values needed to detect the carbon G-band with R$\sim$3000 as a function of \teff\, and A(C).\label{tab_car}}
\begin{tabular}{lllll}
 \hline
 A(C)  & $<5300$\,K  & $5300-5900$\,K &$5900-6450$\,K&$>6450$\,K \\
\hline
 $\geqslant\odot$& 8 &8  & 8 & 12  \\
 $8.4 -7.4$& 8&10 & 25  & 60 \\
  $7.4 -6.4$ & 10 & 25 &  100& 800\\
  $6.4 -5.4$& 15 &100  &--  & -- \\
 $5.4 -4.4$  & 45 & -- & -- & --\\
  $4.4- 3.4$&  900 & -- & -- & -- \\
   $>3.4$   &--     & -- & -- &  -- \\
\hline
\end{tabular}
\end{center}
\end{table}

In order to better understand the systematics involved in the determination of carbon, we assess its correlation with the determination of \logg. In Figure \ref{gravity_plot}, we compare the derived \car\, with those values we find if we fix $\logg=2.5$ as a function of \logg. The points are also coloured according to the \teff. The most relevant part of the plot is the giant regime since this is where the majority of the sample with good \car\, determinations are located, due not only to the lower temperatures (See Table \ref{tab_car}), but also because at these \logg\, values carbon is more likely to be overestimated and therefore considered to be reliably determined based on the criteria in Table \ref{tab_car}. As a result, for the stars for which we derive carbon abundances, we have systematic uncertainties which are large but well delimited, especially at S/N <25. Therefore, we assume systematic uncertainties from 0.2\,dex up to 0.6\,dex, depending on the S/N of the spectrum and subsequent reliability of the \logg.
%The distribution of carbon abundance as a function of metallicity is shown in Figure \ref{carbon_plot2}, including both the A(C)$-$[Fe/H] diagram \citep{spi13,yoo16} and the [C/Fe]$-$[Fe/H] version of the same diagram.

%As discussed at length in \citet[][and references therein]{bee05, yong13II,boni15}, the number of carbon-enhanced metal-poor (CEMP) stars increases with decreasing metallicity. In particular, only one star, J1029+1729, belonging to the \feh$<-4.5$ regime is clearly carbon-normal \citep{caff11} with $[C/Fe]<0.7$. In addition, $Pristine\,221.8781+9.7844$ with $[C/Fe]<1.76$ could potentially also be part of this rare group. However, all of the other 12 most metal-poor stars already known are CEMP stars (86\%). Following \citet{pla14}, 43\% of stars with $-4<\feh<-3$ are indeed CEMP stars while we find 50\% in our \pris sample. For the very metal-poor regime with $-3<\feh<-2$ we find 37\% with carbon enhancement.

\section{Results }\label{catalogue}

\renewcommand{\thefootnote}{\fnsymbol{footnote}}
\begin{table*}
\caption{Metallicities, temperatures, and carbon abundances of the \pris spectroscopic sample. Uncertainties include systematic and statistical errors. We only include a small sample of 9 of the \totstars stars observed, to illustrate the structure of the table. The columns are described in more detail in the text. The full table is available online. The adopted solar abundances are those from \citet{asp05}.}
\begin{center}
\begin{tabular}{lcccccccccccc}
\hline 
\multicolumn{1}{|c|}{\textbf{Name}} &
\multicolumn{1}{c|}{\textbf{$V$\footnotemark[2]}} &
\multicolumn{1}{c|}{\textbf{$CaHK$}} & \multicolumn{1}{c|}{\textbf{[Fe/H]}} &  \multicolumn{1}{c|}{\textbf{[Fe/H]}}   & \multicolumn{1}{c|}{\textbf{Teff}}  & \multicolumn{1}{c|}{\textbf{\logg}}  & \multicolumn{1}{c|}{\textbf{[C/Fe]}}   & \multicolumn{1}{c|}{\textbf{S/N}}  & \multicolumn{1}{c|}{\textbf{Flag}} & 
\multicolumn{1}{c|}{\textbf{Previously}}\\

\multicolumn{1}{|c|}{Origin} & \multicolumn{1}{c|}{SDSS} &
\multicolumn{1}{c|}{Pristine} & \multicolumn{1}{c|}{Pristine} & \multicolumn{1}{c|}{FERRE} & \multicolumn{1}{|c|}{FERRE}  & 
\multicolumn{1}{c|}{FERRE}  & \multicolumn{1}{c|}{FERRE} & \multicolumn{1}{|c|}{FERRE} & \multicolumn{1}{|c|}{\textbf{Q,C}} & 
\multicolumn{1}{c|}{\textbf{observed}} \\

\multicolumn{1}{|c|}{Units} & \multicolumn{1}{c|}{mag} &\multicolumn{1}{c|}{mag} & \multicolumn{1}{c|}{} & \multicolumn{1}{c|}{}   & \multicolumn{1}{c|}{K}   &  \multicolumn{1}{c|}{}  & \multicolumn{1}{c|}{}& \multicolumn{1}{c|}{$\rm px^{-1}$}\\

\hline 
P138.xxxx+16.xxxx\footnotemark[1] & 16.06 & 16.90 & -2.84 & $-3.4\pm 0.2$  & $5266\pm 122$  & $4.8\pm 0.5$  & $0.3\pm 0.4$&  34 & X,-1 & -     \\
P149.1350+15.0447 & 16.06 & 16.38 & -2.72 & $-2.5\pm 0.2$  & $6047\pm 104$  & $1.1\pm 0.5$  & $0.5\pm 1.0$&  49 &  X,-1 & -  \\
P151.4987+13.9300 & 16.67 & 16.90 & -2.84 & $-2.9\pm 0.6$  & $5380\pm 176$  & $1.0\pm 0.5$  & $1.0\pm 0.5$&  26 &  X,-1 & LAMOST,SEGUE   \\
\hline                                                                                                                           
P184.xxxx+43.xxxx\footnotemark[1] & 15.92 & 16.71 & -2.94 & $-3.7\pm 0.2$  & $5509\pm 103$  & $4.9\pm 0.5$  & $0.4\pm 0.5$&  24 &  X,-1 & - \\
P185.8616+41.3093 & 15.83 & 16.71 & -2.87 & $-2.8\pm 0.2$  & $5264\pm 104$  & $1.4\pm 0.6$  & $0.8\pm 0.4$&  25 &  X,1 & SEGUE  \\
P218.6977+15.5932 & 15.57 & 15.89 & -2.93 & $-3.1\pm 0.2$  & $6305\pm 111$  & $4.4\pm 0.5$  & $0.7\pm 0.8$&  28 &  X,-1 & -  \\
\hline                                                                                                                           
P220.7009+13.1405 & 16.88 & 17.49 & -3.40 & $-3.3\pm 0.2$  & $5464\pm 111$  & $3.5\pm 0.5$  & $1.1\pm 0.4$&  26 &  X,-1 & -  \\
P235.0067+07.1438 & 16.88 & 17.12 & -3.10 & $-3.0\pm 0.2$  & $6216\pm 108$  & $4.9\pm 0.5$  & $0.0\pm 0.7$&  27 &  X,-1 & SDSS   \\
P256.0374+17.0031 & 16.51 & 17.12 & -2.97 & $-2.6\pm 0.2$  & $5552\pm 117$  & $3.6\pm 0.5$  & $0.3\pm 0.2$&  45 &  X,-1 & -  \\
\hline
\end{tabular}
\begin{flushleft}
\footnotemark[1]\textbf{Coordinates of select stars have been removed as they are the subject of an ongoing high-resolution follow-up study (Kietly et al. in prep.)}\\
\footnotemark[2]\textbf{derived using SDSS g and r according to Lupton (2005), \url{https://www.sdss3.org/dr8/algorithms/sdssUBVRITransform.php}}
\end{flushleft}
\label{full_table}
\end{center}
\end{table*}

\subsection{Comparison of the photometric metallicities and spectroscopic metallicities}
\label{photometric_metallicites}

Photometric metallicities were derived using the \pris narrow-band photometry and the SDSS broad-band photometry. The detailed methods of this procedure are described in \cite{sta17I}.

\begin{figure*}
	\includegraphics[width=\textwidth]{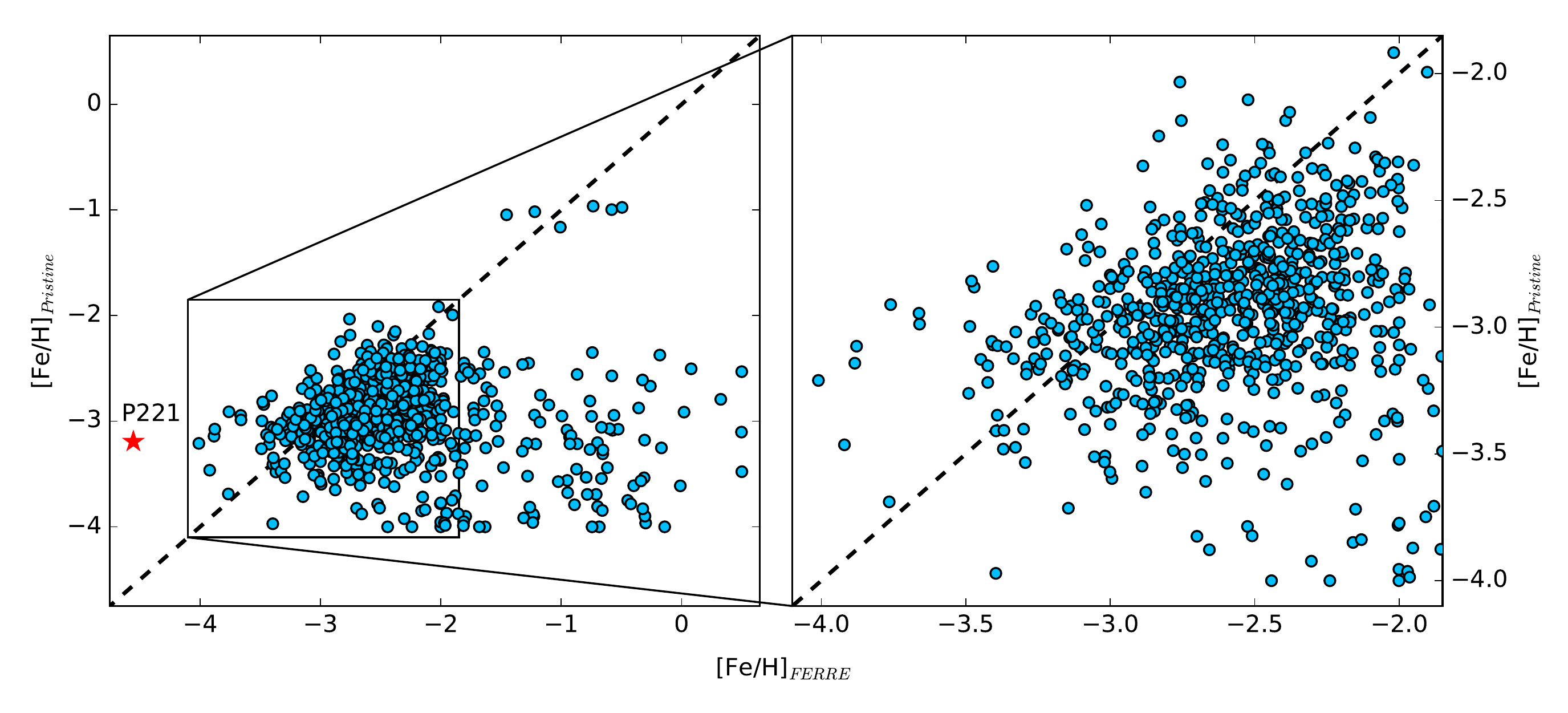}
        \caption{ Photometric metallicities derived with \pris  (\feh$_{Pristine}$) versus spectroscopic metallicities derived with FERRE (\feh$_{FERRE}$) for the total sample. The red star represents $Pristine\,221.8781+9.7844$ from \citet{sta18}. The left panel shows the full metallicity space covered by the follow-up sample, and the right panel shows a zoom in of the highest density region around $-4<$\feh$<-2$.}
    \label{FERRE_vs_Pristine}
\end{figure*}

Figure \ref{FERRE_vs_Pristine} shows the relation between the photometric and spectroscopic metallicities. In the left panel, we show the total parameter space occupied by the data, and in the right panel is a zoomed-in view to better show the details of the plot. We only plot the 863 stars for which there are reliable FERRE and \pris metallicity determinations. For the former, these are stars flagged with ``X" in Table \ref{full_table}, described in Section \ref{the_full_sample}. For the later, we have removed stars that exhibit variability, that may be white dwarfs, that are identified as non-point sources in their PSFs, and that are flagged as being problematic in their SDSS $g$ or $i$ broad band magnitudes (mainly bright sources that show some saturation). These criteria are described in greater detail in the list below Figure 3 in in KY17. Here, we have omitted all criteria based on metallicity but keep all criteria pertaining to photometric quality. Many of the removed stars were observed early on in the follow-up campaign, as we were improving our selection of targets. They are, however, still included in the full table for completeness since the derived spectroscopic metallicities are not affected by the problematic photometry.
    
In Figure \ref{FERRE_vs_Pristine}, most of the stars are clustered at $-3.5<\feh<-2.0$ due to our follow-up strategy of the best metal-poor candidates first. Since there are more metal-rich stars than metal-poor stars, the metal-rich stars will scatter into the metal-poor regime with a higher frequency than the other way around, and the relative contamination will be higher at the metal-poor end. The combination of this effect and the photometric selection function from the follow-up strategy produces the offset from the 1 to 1 relation (black-dotted line). The right panel also shows a fairly significant dispersion, but given that the uncertainties are on the order of $\sim$ 0.2 dex for both the vertical and horizontal axes, it is not surprising to see a dispersion of $\sim$0.5 dex, although the scatter seems more severe due to the small range in metallicities covered by the data, and the outliers at [Fe/H]$_{FERRE} < -2$. We also note that it is not crucial to have a tight relation in this space, because a coarse differentiation of stars as EMP or VMP is enough to identify promising candidates for follow-up, as well as for much of the interesting ancillary science cases.

There is a distinct population of stars for which the photometric metallicities from \pris are highly discrepant from the spectroscopic metallicities. These are seen in the left panel of Figure \ref{FERRE_vs_Pristine}, as the tail of stars extending to $\feh_{FERRE} > -2$. The criteria for selecting stars for spectroscopic follow-up was investigated and summarized in detail in KY17. Despite ensuring good quality photometry, cleaning white dwarfs (cutting all stars with $(u_0-g_0) < 0.6$) and variable stars, there are still 12\% of stars predicted to have $\feh_{Pristine} < -2.5$ that have $\feh_{FERRE} > -2$. This number rises to 18\% for $\feh_{Pristine} < -3$ (see Table \ref{success_rates_table}). Many of these stars have a large temperature discrepancy between spectroscopy and photometry (|$\Delta$\teff|\,> 500 K for $\sim$ 40\% of these stars), which probably indicates problems with the SDSS broad-band photometry for these stars. This would, in turn, affect the colour, and thus the measured photometric metallicity. In addition, some of this contamination may be attributable to long-period variable stars that were not detected in the Pan-STARRS1 variability catalogue, non-stellar objects, or chromospherically active stars with Ca H \& K in emission (although we note that only 9 such objects with peculiar spectra were identified in the follow-up spectroscopy).

At the lowest metallicities of $\feh_{Pristine} < -3.5$, the percentage of stars with spectroscopic $\feh > -2$ rises to 57\%. This clearly indicates an increasing contamination fraction with decreasing metallicity. Although the slope in this region of the Milky Way metallicity distribution function is not well constrained, it is known to be quite steep, such that stars at these metallicities are incredibly rare with respect to stars of higher metallicity. As a result, even a small number of interloping higher metallicity stars can dominate the candidate sample at these low metallicities.

%Some of this may be attributable to errors in the spectroscopic determination The median S/N for the entire sample is $\sim 23$, whereas for these stars it is $\sim 19$.
 
\subsection{Updated purity and success rates of the \pris survey}
\label{success_rates}
The success rates of the \pris survey were reported after the first year of spectroscopic follow-up using a sample of 205 stars observed at medium-resolution at the WHT and INT (KY17). Due to the small size of that sample, the success rates for finding metal-poor stars computed from them were preliminary estimates. Now that we have a larger follow-up sample of nearly 5 times as many stars, we can update these numbers with better statistics. In order to remain consistent and to allow for an easy comparison, we will use the same metrics to quantify the purity and success rates as were used in KY17, namely: 

\begin{equation*}
\mathrm{success\, rate\,} \% = \frac{\feh_{FERRE} < X}{\feh_\pris < X} \times 100,
\end{equation*}

where FERRE refers to the spectroscopically derived [Fe/H], \pris to the photometric prediction by \pris and X the metallicity limit of interest.

For all of the stars included in Table \ref{full_table}, we did not make a cut in S/N, but rather checked by eye the goodness of the fit for the synthetic spectrum by FERRE. The reason for this is because stars that are cooler and more metal-rich have larger absorption lines, and are therefore easier to identify at lower S/N than stars that are hotter and more metal-poor. As a result, we successfully determine the metallicities for more stars at higher metallicities ([Fe/H] > -2) with low S/N values (S/N < 15), rather than to cut these stars out with a S/N cut. However, for the calculation of the success rates, this would bias our sample with more metal-rich stars and fewer metal-poor stars. Therefore, we compute the success rates using only stars with S/N > 25, the regime in which we can reliably measure metallicities, even at [Fe/H] < -3. Taking this sample, we find a success rate of \SRemp\% for finding stars with \feh < -3.0, and \SRvmp\% for finding stars with \feh < -2.5. In KY17, we reported a success rate of 22\% for \feh < -3.0, and 70\% for \feh < -2.5. This discrepancy can be attributed to the cut at S/N > 25. If we make the same cut in the KY17 sample, this decreases the success rates to 20\% and 58\% for \feh < -3 and \feh < -2.5, respectively, meaning that these values are fully compatible with what we find in this work. In KY17, we did not originally make a cut at S/N > 25 when computing the success rates as this would have reduced the sample from 205 down to 62 stars, leading to uncertainties of low number statistics. In the current work, making this cut still leaves 331 stars, and still allows for a robust determination of the success rates. 

We therefore update the success rates of the \pris survey to \SRemp\% for \feh < -3.0, and \SRvmp\% for \feh < -2.5. These values, along with other diagnostics, such as the contamination rate (fraction of stars with \feh\, > -2) are summarized in Table \ref{success_rates_table}.

\subsection{The carbon-enhancement present in the sample}

\begin{figure}	
\begin{center}	
	\includegraphics[width=80mm]{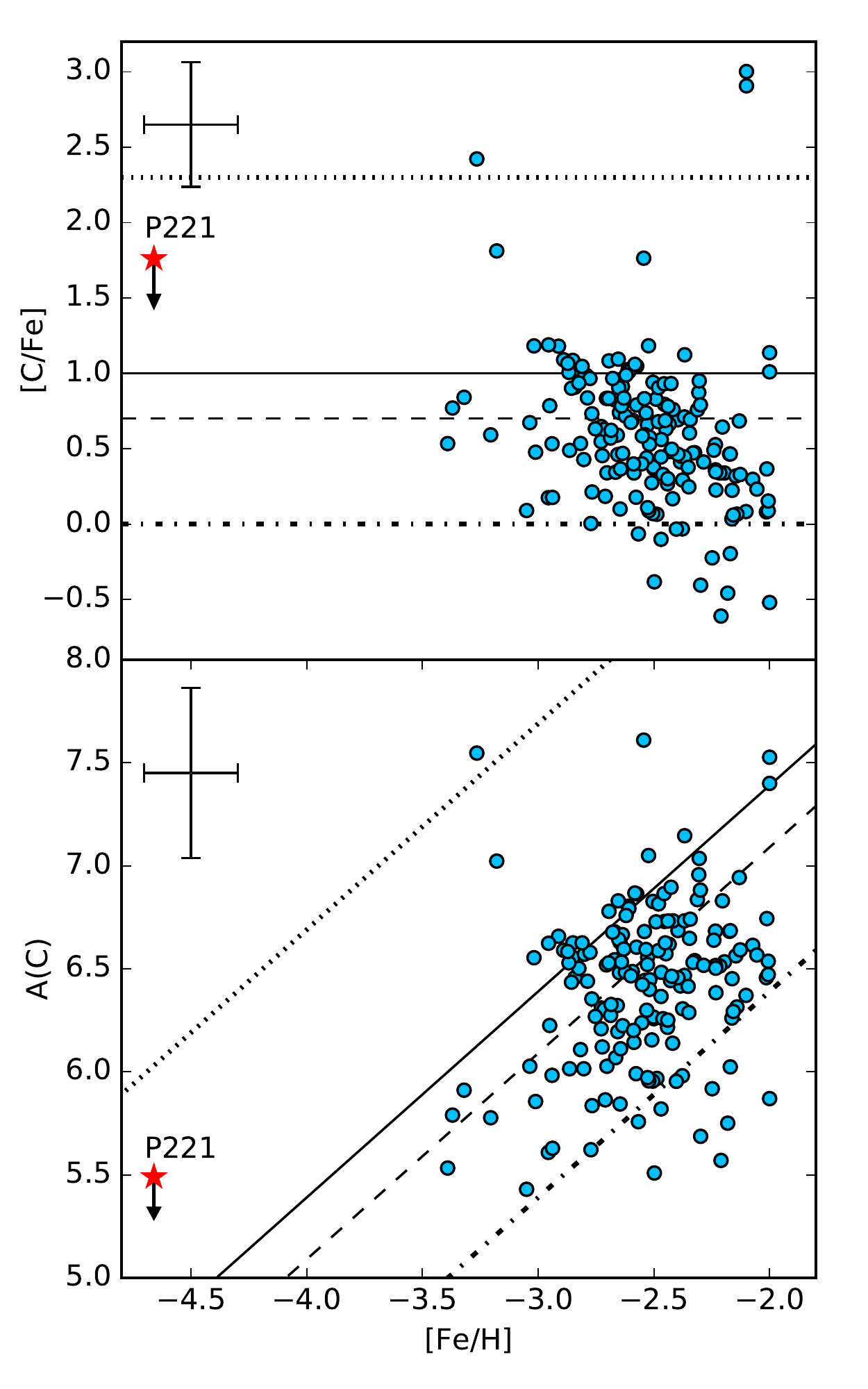}
        \caption{Carbon vs iron for [C/Fe] (top) and absolute carbon (bottom). The red star represents $Pristine\,221.8781+9.7844$ with parameters derived from the analysis of a high-resolution UVES spectrum \citep{sta18}. The dashed-dotted line at [C/Fe] = 0 shows the solar carbon abundance, the dashed and solid lines at [C/Fe] = +0.7 and +1.0 show the thresholds for carbon enhancement defined in \citet{aoki07} and \citet{bee05}, respectively, and the dotted line at [C/Fe] = 2.3 shows the boundary of the silicate dominated region, as described in \citet{chi17}. The error bars shown in the top left of each panel represent the median errors of the sample.}
    \label{carbon_plot2}
    \end{center}
\end{figure}

Figure \ref{carbon_plot2} shows the distribution of absolute carbon (A(C); bottom panel) and [C/Fe] abundances as a function of metallicity for the 169 stars for which we are able to make a reliable carbon determination (Section \ref{carbon_sec}). Both CEMP reference lines at [C/Fe] = 1.0 \citep{bee05} and [C/Fe] = +0.7 \citep{aoki07} are plotted as solid and dashed lines, respectively. The high-resolution carbon abundance value for $Pristine\,221.8781+9.7844$ is also included as the red star. 

To compute the CEMP fractions, we first draw a new sample of values for the [C/Fe], and [Fe/H] measurements, taking into account both the statistical and systematic uncertainties of each. We then compute the fraction of stars with [C/Fe] above the two limits of [C/Fe] = +1.0 and +0.7, and repeat this exercise $10^6$ times in a Monte Carlo fashion. The resulting distributions are approximately Gaussian in shape, and are therefore reasonably well described by a mean and standard deviation. For the \feh\, < -3 sample, we compute CEMP fractions of 58 $\pm$ 14\% and 43 $\pm$ 13\% for the [C/Fe] > +0.7 and +1.0, respectively. For the -3 < \feh\, < -2 sample, we compute CEMP fractions of 41 $\pm$ 4\% and 23 $\pm$ 3\% for [C/Fe] > +0.7 and +1.0, respectively. 

\citet{pla14} find 43\% of stars to have $\car \geq +0.7$ and -4 < \feh < -3, a value which differs at the 1$\sigma$ level compared to the 58 $\pm$ 14\% derived in this work, and is therefore not statistically significant. Furthermore, \citet{Norr19} perform a rigorous analysis of the 3D and NLTE corrections relevant for the carbon abundance determinations, and demonstrate a significant decrease in the carbon content for a number of CEMP stars from the literature when full 3D-NLTE corrections are taken into account. The CEMP-no group are stars that do not show significant enrichment in neutron-capture elements (s- and r-process), and are the most numerous subgroup among CEMP stars. As a result of those 3D and NLTE corrections in \citep{Norr19} a significant number of CEMP-no stars become carbon-normal. However, we do not know the fraction CEMP-no stars in our current sample, but if we consider that a similar fraction of them likely are, as is the case in the literature, it is likely that the computed CEMP fractions would decrease considerably. It is therefore difficult to draw firm conclusions from this current sample of CEMP stars, but further, more detailed follow-up --particularly targeting carbon and the neutron capture elements in the EMP stars-- could potentially be a very nice sample with which to investigate this further. 

%This may require a quantitative and qualitative revision of the role of the carbon-enhancement in the formation of metal-poor stars, since the relative fraction seems to be lower than what has recently been reported in the literature.

%\begin{figure}	
%\begin{center}	
%	\includegraphics[width=80mm]{CEMP_MC_kernel_1.jpg}
%        \caption{CEMP fraction }
%    \label{carbon_kernel}
%    \end{center}
%\end{figure}

\subsection{The full sample}
\label{the_full_sample}
In this paper we present a full catalogue from three years of follow-up spectroscopy of \pris candidates. The full table, consisting of \totstars stars is available online. An abbreviated version of the full table showing the provided columns as well as a sample of 9 rows is shown in Table \ref{full_table}. The column $CaHK$ is the magnitude obtained from the \pris narrow-band filter, the column $\textrm{[Fe/H]}_{Pristine}$ is the photometric metallicity determined using the $(g-i)_{0}$ SDSS colours and \textit{Pristine} photometry (described in \citet[][Section 3.2]{sta17I}). The next two columns are the spectroscopic metallicities, effective temperatures and surface gravities derived from FERRE and their associated uncertainties. Column S/N is the signal-to-noise ratio of the analyzed spectrum. We also provide a Q-flag, representing the reliability of the spectroscopic metallicity determination. An entry of "X" indicates that the synthetic spectral fit was reliable and that the given \feh$_{\rm FERRE}$ value can be trusted to within the provided uncertainties (93\% of the sample have this flag). In order to provide as much information as possible, we also provide tentative metallicity values for stars for which the S/N is too low for a robust determination of stellar parameters, but that still have some information in the observed spectrum. These stars are given a flag of T (6\% of the sample), and are good candidates to be re-observed with higher S/N and at higher resolution facilities. The C-flag shows if the carbon determination is reliable (value $1$) or not (value $-1$), and was derived based on S/N and temperature criteria described in Section \ref{carbon_sec}. The last column indicates whether the object was already spectroscopically observed by other surveys. Finally, the object coordinates are contained in the name, but we provide these explicitly as RA and Dec in degrees on the online version of the table.

There are a small number of stars for which the \pris metallicity classification fails, meaning that from photometry the object was expected to be a metal-poor star, but from spectroscopy it was determined to be some other type of object. These could be stars with CaHK in emission, non-stellar objects, or various other objects with unusual behaviour in the CaHK region. However, this only occurs for 9 of the observed objects ($< 1\%$ of the sample), indicating that the sample is well cleaned. We remove these 9 objects from the catalogue since both their photometric and spectroscopic metallicities are unreliable, but consider them in the sample when computing the success rates since they do contribute to the contamination.

\begin{table*}
\centering
\caption{Numbers of stars with photometric predictions $\textrm{\rm [Fe/H]}_{\textrm{Pristine}}$ below $-2.5$ and $-3.0$, the numbers of stars that are spectroscopically confirmed below those metallicities, and the success rates, given for all stars with S/N > 25, with the selection criteria applied (described partially in Section \ref{photometric_metallicites}, and in detail in KY17), and the sample of stars with $\textrm{\rm [Fe/H]}_{\textrm{Pristine}} \leq -3.0$.}

\label{success_rates_table}
\begin{tabular}{cccc}
\hline
\textbf{}                   & \textbf{All stars} & \textbf{Selection criteria} & \textbf{$\mathrm{[Fe/H]}_{Pristine} \leq -3.0$} \\
\textbf{}                   & \textbf{S/N > 25} & \textbf{S/N > 25} & \textbf{S/N > 25}
\\ \hline
\textbf{Total number} & 344 & 331 & 129 \\
\textbf{$\textrm{[Fe/H]}_{\textrm{Pristine}} \leq -2.5$} & 325/344 (94\%) & 315/331 (95\%) & 129/129 (100\%) \\
\textbf{$\textrm{[Fe/H]}_{\textrm{Pristine}} \leq -3.0$} & 132/344 (38\%) & 129/331 (39\%) & 129/129 (100\%) \\
\hline
\textbf{$\textrm{[Fe/H]}_{\textrm{FERRE}} \leq -2.5$} & 184/344 (53\%) & 180/331 (54\%) & 76/129 (59\%) \\
\textbf{$\textrm{[Fe/H]}_{\textrm{FERRE}} \leq -3.0$} & 48/344 (14\%) & 47/331 (14\%) & 30/129 (23\%) \\
\hline
\textbf{$\textrm{[Fe/H]}_{\textrm{FERRE}} \geq -2.0$} & 45/344 (13\%) & 39/331 (12\%) & 23/129 (18\%) \\
\hline
\textbf{success $\textrm{[Fe/H]} \leq -2.5$} & 178/325 (55\%) & \textbf{175/315 (56\%)} & - \\
\textbf{success $\textrm{[Fe/H]} \leq -3.0$} & 30/132 (23\%) & 30/129 (23\%) & \textbf{30/129 (23\%)} \\
\hline
\end{tabular}

\end{table*}

\begin{table}
\centering
\caption{Number of candidate stars in different magnitude bins and metallicity ranges. The first number in each cell is the number of stars followed up with spectroscopy from the sample in this paper, and the second  is the total number of candidates as of the time of publication over the $\sim 2500 \textrm{ deg}^2$ of the \pris footprint used to select candidates that are the focus of this paper. [Fe/H] values shown are photometric \pris metallicities.}
\label{Spectro_sample}
\begin{tabular}{cccc}
\hline
\textbf{}                   & \textbf{\# Candidates} & \textbf{{[}Fe/H{]}$\leq$-2.5} & \textbf{{[}Fe/H{]}$\leq$-3.0} \\ \hline
\textbf{$V < 15$}             & 169/509           & 139/293                           & 66/92                              \\
\textbf{$15 < V < 16$} & 536/1 809          & 475/989                            & 160/206                              \\
\textbf{$16 <V < 17$} & 246/5 423         & 238/2 785                           & 148/540                            \\
\textbf{$17 < V < 18$} & 57/14 682          & 56/7 321                           & 43/1 393                             \\ 
\textbf{$18 < V < 19$} & 0/35 036          & 0/16 887                           & 0/3 977                             \\ \hline
\textbf{Total }                      & 1008/57 459        & 908/28 275                             & 417/6 208                            \\
\hline
\end{tabular}
\end{table}

\section{Future of the survey}\label{future}
In addition to hunting for the most metal-poor stars in the Galaxy, the photometric metallicities that are produced by the narrow-band photometry of the \pris survey can be used for several other interesting science cases. For instance, \citet{long18} conducted an in depth study of the metallicity distribution and velocity dispersion of the faint Milky Way satellite \textit{Draco II} using $CaHK$ photometry, and work is ongoing on a similar analysis to characterize the properties of many other nearby satellites \citep{long19}. Another study by Starkenburg et al. (2019, subm.) demonstrated the powerful capabilities of the \pris narrow-band filter to identify blue horizontal branch (BHB) stars and disentangle them from the contaminating blue straggler (BS) population, providing a uniquely clean sample of distance indicators with which to study the outer reaches of the Galactic halo. 
Finally, Arentsen et al. (2019, in prep.) are studying the metal-poor component of the Galactic bulge with the Pristine Inner Galaxy Survey (PIGS).

%A study by Youakim et al. in prep is using the \pris metallicities to chemically separate the Galactic halo and analyze the substructrue content as a function of metallicity.

\subsection{Pristine and Gaia}
The highly anticipated Gaia data have initiated a revolution in the study of galactic archaeology and it is changing our understanding of the Galaxy \citep{gaia2018}. The latest data release provided for high-precision astrometry measurements and 3-filter photometry for over 1.3 billion. The range of possibilities for using Gaia photometry together with more than 5 million \pris metallicity determinations are broad, and open the door to an unprecedented mapping of the Galaxy using the full six-dimensional phase-space plus metallicity information. For example, work is ongoing using Gaia and \pris to study the substructures present in and around our Galaxy and their dependence on metallicity, as well as an analysis of the metallicity distribution function of the halo at the lowest metallicities (Youakim et al. in prep.). On the other hand  exquisite Gaia parallaxes, proper motions and photometry allow us to derive surface gravities, effective temperatures and orbits for EMP stars \cite[see, e.g.,][]{boni19,fre18,ses18}. The dynamics of the most ancient stars of the Milky Way could be a crucial piece of information for understanding the formation and evolution of the Galactic halo. For example, recent work by \citet{ses18}, demonstrated that an important fraction of the known UMP stars seem to have orbits that are confined to Galactic plane, suggesting interesting new scenarios for their origins. In addition, a complete kinematical analysis of the sample presented in this paper will be presented in Sestito et al. (2019, in prep.). Finally, \citet{boni19II} combined Gaia parallaxes and \pris photometry to  derive photometric metallicities, effective temperatures and surface gravities. These authors also studied the chemical composition and ages of 40 metal-poor stars with the SOPHIE high-resolution spectrograph.

\subsection{Pristine and WEAVE}
%The impending arrival of the new large spectroscopic surveys will nicely complement the still ongoing Gaia project. A new, deeper view -- not only kinematically but also chemically -- of the Milky Way halo, will shed light on the formation and evolution of the Galaxy. This unprecedented amount of high quality data will greatly expand the capabilities of the Galactic archaeology community thanks to surveys like 4MOST, DESI or WEAVE.  The successful ratios presented in Table \ref{success_rates} demonstrate that the \pris filter is one of the best ways to pre-select EMP candidates to observe in those surveys. In particular, if the WEAVE project would devote few tens of fibers per pointing in the Galactic archaeology sub-survey, it would be possible to increase the number of EMP and UMP stars with studied chemical signature by one order of magnitude. This means that after five years of observing a $\sim$6000\,$\rm deg^{2}$ footprint we could measure chemical abundances such as C, Na, Mg, Al, Si, Ca, Ti, Fe, for about $\sim$ 3000 stars with \feh < -3.0 and $\sim150/200$ stars with \feh < -4.0, including 6/11 hyper metal-poor stars ([Fe/H]<-5.0), doubling the samples currently available from several decades of efforts.
The impending arrival of the new large spectroscopic surveys will nicely complement the still ongoing Gaia project. A new, deeper view --not only kinematically but also chemically-- of the Milky Way halo, will shed light on the formation and evolution of the Galaxy. This unprecedented amount of high quality data will greatly expand the capabilities of the Galactic archaeology community thanks to surveys like 4MOST, DESI or WEAVE. The success rates presented in Table \ref{success_rates_table} demonstrate that the Pristine filter is one of the best ways to pre-select EMP candidates to observe in those surveys. In particular, the WEAVE project will devote up to twenty fibers per WEAVE 3.14 deg$^2$ field of view to Pristine-selected EMP candidates in the magnitude range $15<G<19$, in the low-resolution Galactic archaeology survey of high Galactic latitudes (Jin et al. 2019 in preparation). Over the planned $\sim$8,500\,deg$^2$ of the survey, of which we anticipate $\geq$5,000 deg$^2$ will be in common with the Pristine footprint at the time they are observed in WEAVE, this adds up to up to $\sim$30,000 candidate EMP stars, of which according to Table \ref{success_rates_table}, $\sim 5,000 - 7,000$ would turn out to be [Fe/H]$<-3.$. This would increase the number of spectroscopically confirmed EMP and UMP stars with known chemical signatures by one order of magnitude. After five years of observing we we expect to have measured the chemical abundances such as C, Na, Mg, Al, Si, Ca, Ti, Fe, for about $\sim3000$ stars with $\feh<-3.0$ and $\sim 150-200$ stars with $\feh<-4.0$, including $\sim 5-10$ hyper metal-poor stars ($\feh<-5.0$), doubling the samples currently available from several decades of efforts.
 Additionally, WEAVE Galactic archaeology high resolution survey (HR) will be able to measure the full suite of chemical signatures for the brightest part of the Pristine sample ($g\leq 15.5$) where it overlaps with the WEAVE HR survey, although the density of such bright targets will be much lower.
 %i.e. $\sim$?? deg$^2$ of the 5600deg$^2$ of the HR survey. The density of such bright EMP stars is much lower, so that the actual yield of this follow-up is expected to include $\sim$7,000 candidate VMP stars. 
\section{Conclusions}\label{conclusions}
Expanding upon the previous work conducted in \citet{sta17I} and KY17, we have presented a sample consisting of 1008 stars, representing three years of follow-up of medium- and low-resolution spectroscopy of EMP candidates from the \pris survey. The number of stars followed-up spectroscopically has increased by a factor of 5, allowing for the success rate of stars with \feh < -2.5 and \feh < -3.0 to be updated to \SRvmp\% and \SRemp\%, respectively. This is a relevant milestone in the field of Galactic archaeology, demonstrating the utility of the \pris filter to select EMP candidates for the next generation of spectroscopic surveys such as WEAVE.
The recent discovery of $Pristine\,221.8781+9.7844$ \citep{sta18}, the second most metal-poor star yet discovered, shows that \pris photometry is also effective in finding UMP stars in the most interesting and poorly populated regime of \feh$<-4.5$.
In addition, we demonstrated that the FERRE code is capable of deriving stellar parameters even at relatively low-resolution, namely with the stars observed with EFOSC2. Furthermore, we show for the first time in the \pris project that we are able to provide individual carbon abundances from measurements of the G-band with moderate S/N in medium-resolution spectra for 169 stars, or $\sim$ 20\% of the total sample, although lower average S/N as compared to DA17 results in higher overall uncertainties in the carbon measurements than previously achieved. With this medium-resolution follow-up spectroscopy sample (along with the previous analysis of KY17), we have been able to thoroughly characterize the photometric selection and success rates of the \pris survey in this magnitude range, and future follow-up is planned to mostly be done with MOS facilities such as WEAVE. More observations with low- and medium-resolution spectroscopic facilities of metal-poor candidates selected from \pris are highly desirable with the aim of increasing the number of ultra/hyper metal-poor stars, but also to provide a larger sample of CEMP and carbon-normal EMP stars.

% used for referring to this section from elsewhere

\section*{Acknowledgements}
 We gratefully acknowledge the Isaac Newton Group (ING) staff, in particular the support astronomers and staff at the INT/WHT for their expertise and help with observations.
 We also thank the staff at ESO for helping during EFOSC observations, and the CFHT staff for performing the observations in queue mode. DA thanks the Leverhulme Trust for financial support. DA acknowledges the Spanish Ministry of Economy and Competitiveness (MINECO) for the financial support received in the form of a Severo-Ochoa PhD fellowship, within the Severo-Ochoa International PhD Program. DA, CAP, and JIGH and CAP also acknowledge the Spanish ministry project MINECO AYA2017-86389-P. JIGH acknowledges financial support from the Spanish Ministry of Science, Innovation and Universities (MICIU) under the 2013 Ram\'on y Cajal program MICIU RYC-2013-14875, and also from the Spanish ministry project MICIU AYA2017-86389-P. ES, KY and AA gratefully acknowledge funding by the Emmy Noether program from the Deutsche Forschungsgemeinschaft (DFG). This work has been published under the framework of the IdEx Unistra and benefits from a funding from the state managed by the French National Research Agency as part of the investments for the future program. NFM, RI, NL, PB, EC, VH, CK, and PS gratefully acknowledge support from the French National Research Agency (ANR) funded project ``Pristine'' (ANR-18-CE31-0017) along with funding from CNRS/INSU through the Programme National Galaxies et Cosmologie and through the CNRS grant PICS07708. The authors benefited from the International Space Science Institute (ISSI) in Bern, CH, thanks to the funding of the Teams ``The Formation and Evolution of the Galactic Halo'' and ``Pristine''.
 The French co-authors acknowledge support from the Agence National de la Recherche (ANR), through contract N. 183787.
CL acknowledges financial support from the Swiss National Science Foundation (Ambizione grant PZ00P2\_168065).
DA thanks F\'atima Mesa-Herrera from Laboratory of Membrane Physiology and Biophysics, University of La Laguna, for those beautiful nights observing the sky at La Palma in December 2017.
We thank the reviewer, Tim Beers, for his thorough review and highly appreciate the comments and suggestions, which significantly contributed to improving the quality of the publication and the definitive shape of the online material.

%%%%%%%%%%%%%%%%%%%%%%%%%%%%%%%%%%%%%%%%%%%%%%%%%%



\bibliographystyle{mnras}

\bibliography{biblio}

%%%%%%%%%%%%%%%%%%%%%%%%%%%%%%%%%%%%%%%%%%%%%%%%%%

%%%%%%%%%%%%%%%%% APPENDICES %%%%%%%%%%%%%%%%%%%%%

% Don't change these lines
\bsp	% typesetting comment
\label{lastpage}
\end{document}